\documentclass{ws-procs9x6-cpt19}

\usepackage[retain-unity-mantissa=false,range-units=single,range-phrase={ -- }]{
siunitx }

\begin{document}

\newcommand{\refeq}[1]{(\ref{#1})}
\def\etal {{\it et al.}}
\newcommand{\bvec}[1]{\boldsymbol{#1}}
%any other macros go here 

\title{Searches for Lorentz-Violating Signals\\with Astrophysical Polarization 
Measurements}

\author{F.\ Kislat}

\address{Department of Physics and Astronomy and Space Science Center,\\
University of New Hampshire,
Durham, NH 03824, USA}

\begin{abstract}
Astrophysical observations are a powerful tool 
to constrain effects of Lorentz-invariance violation in the photon sector.
Objects at high redshifts provide the longest possible baselines, 
and gamma-ray telescopes allow us to observe some of the highest energy photons.
Observations include polarization measurements 
and time-of-flight measurements of transient or variable objects 
to constrain vacuum birefringence and dispersion.
Observing multiple sources covering the entire sky 
allows the extraction of constraints on anisotropy.
In this paper, 
I review methods and recent results 
on Lorentz- and CPT-invariance violation constraints 
derived from astrophysical polarization measurements 
in the framework of the Standard-Model Extension.
\end{abstract}

\bodymatter

\section{Introduction}
Observer-independence of the speed of light 
postulated by Einstein's Theory of Special Relativity 
implies that the speed of light for a given observer 
is independent of frequency, 
direction of propagation, 
and polarization.
Conversely, 
if Lorentz symmetry is broken, 
any of these may no longer hold, 
resulting in a vacuum 
that effectively acts like an anisotropic, birefringent, and/or dispersive medium.
Theories of quantum gravity suggest 
that Lorentz invariance may be violated at the Planck scale, 
but effects must be strongly suppressed at attainable energies.
Photons propagating over astrophysical distances 
enable some of the strongest tests 
as tiny effects will accumulate during propagation of the photons.

The Standard-Model Extension\cite{sme,sme_electro_2009} (SME) 
is an effective-field-theory approach 
to describe low-energy effects of a more fundamental high-energy theory.
It allows a categorization of effects, 
described in essence by a set of coefficients 
that are characterized in part by the mass dimension $d$ 
of the corresponding operator.
The SME photon dispersion relation can be written as
\begin{equation}
  E = \left(1 - \varsigma^0 \pm \sqrt{(\varsigma^1)^2 + (\varsigma^2)^2 +
                                      (\varsigma^3)^2}\right) p,
\end{equation}
where the $\varsigma^i$ can be written 
as an expansion in $d$ and spherical harmonics.
This results in $(d-1)^2$ independent coefficients $k_{(V)jm}^{(d)}$ for odd $d$ 
describing anisotropy, dispersion, and birefringence; 
for even $d$, 
$2(d-1)^2-8$ independent birefringent coefficients 
$k_{(E)jm}^{(d)}$ and $k_{(B)jm}^{(d)}$ 
as well as $(d-1)^2$ nonbirefringent coefficients 
$c_{(I)jm}^{(d)}$ emerge.
Effects of mass dimension $d$ 
are typically assumed to be suppressed by $M_\text{Planck}^{d-4}$.

%where the $\varsigma^i$ can be written 
%as an expansion in $d$ and spherical harmonics.
%This results in $(d-1)^2$ independent coefficients $k_{(V)jm}^{(d)}$ for odd $d$ 
%describing anisotropy, dispersion, and birefringence; 
%as well as, for even $d$, 
%$2(d-1)^2-8$ independent birefringent coefficients $k_{(E)jm}^{(d)}$ and 
%$k_{(B)jm}^{(d)}$ and $(d-1)^2$ non-birefringent coefficients 
%$c_{(I)jm}^{(d)}$.
%Effects of mass dimension $d$ are typically assumed to be suppressed by 
%$M_\text{Planck}^{d-4}$.

The individual coefficients 
$k_{(E)jm}^{(4)}$, $k_{(B)jm}^{(4)}$, $k_{(V)jm}^{(5)}$, and $c_{(I)jm}^{(6)}$ 
have been constrained 
using astrophysical polarization\cite{kislat_krawczynski_2017,kislat_2018} 
and time-of-flight measurements,\cite{kislat_krawczynski_2015} 
and the $k_{(V)jm}^{(3)}$ have been constrained 
using CMB measurements.\cite{cmb} 
At $d > 6$, 
linear combinations of coefficients have been constrained.\cite{datatables} 
Here, 
we will review some recent results 
obtained from polarization measurements of astrophysical objects.

\section{Birefringence}
The polarization of an electromagnetic wave 
can be described by the Stokes vector $\bvec{s} = (q, u, v)^T$, 
where $q$ and $u$ describe linear polarization 
and $v$ describes circular polarization.
Vacuum birefringence results in a change of the Stokes vector 
during the propagation of a photon given by\cite{sme_electro_2009}
\begin{equation}\label{eq:dsdt}
  \frac{d\bvec{s}}{dt} = 2E\bvec{\varsigma} \times \bvec{s},
\end{equation}
which represents a rotation of $\bvec{s}$ along a cone 
around the birefringence axis 
$\bvec{\varsigma} = (\varsigma^1, \varsigma^2, \varsigma^3)^T$.

Most astrophysical emission mechanisms 
do not result in a strong circular polarization, 
but linear polarization can be fairly large 
reaching multiples of ten percent.
In general, 
Eq.~\eqref{eq:dsdt} results in a change of the linear polarization angle (PA) 
as the photon propagates, 
as well as linear polarization partially turning into circular polarization.
However, 
SME operators of odd and even $d$ 
result in distinctly different signatures.

If only odd-$d$ coefficients are nonzero, 
$\varsigma^1 = \varsigma^2 = 0$ 
and $\bvec{\varsigma}$ is oriented along the circular-polarization axis $\bvec{v}$.
In this case, 
linear polarization remains linear, 
but the PA makes full \ang{180} rotations 
as $\bvec{s}$ will rotate in the $q$--$u$ plane.
Any circular polarization would remain unaffected.
If only even-$d$ coefficients are nonzero, 
$\varsigma^3 = 0$ and 
$\bvec\varsigma$ is in the $q$--$u$ plane.
Then, 
a Stokes vector with an initial $v = 0$ 
will in general rotate out of this plane 
acquiring a circular polarization, 
while at the same time the PA will undergo swings 
with a magnitude depending on 
the angle between $\bvec{s}$ and $\bvec\varsigma$.
However, 
in this case linear polarization 
with a polarization angle of 
$\text{PA}_0 = \frac{1}{2}\arctan\left(-\varsigma^2/\varsigma^1\right)$ 
will remain unaffected.

For all $d \geq 4$, 
Eq.~\eqref{eq:dsdt} results in an energy-dependent rate of change of $\bvec{s}$.
Hence, 
birefringent Lorentz-violating effects 
can be constrained by measuring the 
(linear) polarization of astrophysical objects as a function of energy.
Spectropolarimetric observations are ideally suited for this purpose 
as the change of PA with energy is measured directly 
resulting in the strongest constraints.

Constraints can also be derived from broadband measurements 
where polarization is integrated 
over the bandwidth of the instrument.\cite{kostelecky_mewes_2013}
In this case, 
the rotation of the PA 
results in a net reduction of the observed polarization fraction (PF).
Given a set of SME coefficients, 
the maximum observable PF, 
$\text{PF}_\text{max}$, 
can be calculated 
assuming the emission at the source is \SI{100}{\percent} polarized.
Any observed $\text{PF} > \text{PF}_\text{max}$ 
would then rule out this set of SME coefficients.
Constraints obtained in this way 
tend to be significantly weaker than those obtained from spectropolarimetry.

Observations of a single source 
can be used to constrain linear combinations of SME coefficients.
Typically, 
limits are obtained by restricting the analysis to a particular mass dimension $d$.
In the even-$d$ case, 
additional assumptions must be made 
due to the existence of the linearly polarized eigenmode of propagation.
Observations of multiple objects 
can be combined to break all of these degeneracies 
and to obtain constraints on individual coefficients, 
even in the even-$d$ case.

The strongest individual constraints 
result from hard x-ray polarization measurements of gamma-ray bursts (GRBs) 
profiting from the high photon energies and high redshifts.
Typical constraints derived from these measurements 
are on the order of \SI{1e-34}{\per\giga\electronvolt} or better 
in the $d=5$ case.\cite{kostelecky_mewes_2013,grbs}
However, 
the number of bursts whose polarization has been measured 
is limited at this point 
and in particular the early results suffer from large systematic uncertainties.
At least for $d < 6$, 
very strong constraints can also be obtained 
from optical polarization measurements of distant objects, 
such as active galactic nuclei and GRB afterglows.
Systematic uncertainties of these measurements are smaller, 
and a much larger number of results is already available.

In two recent papers,\cite{kislat_krawczynski_2017,kislat_2018} 
we have combined more than 60 optical polarization measurements, 
both spectropolarimetric and broadband integrated measurements, 
to constrain individually all birefringent coefficients 
of mass dimensions $d=4$ and $d=5$.
Constraints on the dimensionless coefficients 
$k_{(E)jm}^{(4)}$ and $k_{(B)jm}^{(4)}$ 
are \num{<3e-34}, 
and the $k_{(V)jm}^{(5)}$ are constrained to \SI{<7e-26}{\per\giga\electronvolt}.

\section{Summary and outlook}
Astrophysical polarization measurements are an extremely powerful tool 
to constrain Lorentz-invariance violation in the photon sector 
due to the extremely long baselines.

In the future, 
new instruments with significantly improved high-energy polarization sensitivity 
will become available.\cite{new_missions}
The IXPE satellite scheduled for launch in early 2021 
will measure x-ray polarization 
in the \mbox{\SIrange{2}{8}{\kilo\electronvolt}} range. 
Compton telescopes, 
such as the balloon-borne COSI or the proposed AMEGO mission concept, 
are sensitive to polarization in the 
\mbox{\SI{500}{\kilo\electronvolt} -- \SI{5}{\mega\electronvolt}} energy range.
A slightly lower energy range is covered by the proposed GRB polarimeter LEAP.
AdEPT is a concept for a gamma-ray pair-production telescope 
that would allow 
measuring the polarization of gamma-rays of tens of \si{\mega\electronvolt}.
The ability to measure polarization at high energies 
is crucial in particular for constraining coefficients of $d \geq 6$.

\end{document}